\documentclass[preprint]{aastex62}

\newcommand\sigsfr{\Sigma_{\rm{SFR}}}
\newcommand\siggas{\Sigma_{\rm{gas}}}

\usepackage{enumerate}
\usepackage{multirow}

\graphicspath{{./}{}}

\received{...}
\revised{...}
\accepted{\today}
\submitjournal{ApJL}

\shorttitle{The Single-Cloud Star Formation Relation}
\shortauthors{R. Pokhrel et al.}

\begin{document}


\title{The Single-Cloud Star Formation Relation} 

\correspondingauthor{Riwaj Pokhrel}
\email{Riwaj.Pokhrel@utoledo.edu, riwajpokhrel@gmail.com}

\author[0000-0002-0557-7349]{Riwaj Pokhrel}
\affil{Ritter Astrophysical Research Center, Department of Physics and Astronomy, University of Toledo, Toledo, OH
43606, USA}

\author{Robert A. Gutermuth}
\affil{Department of Astronomy, The University of Massachusetts, Amherst, MA, 01003, USA}

\author{Mark R. Krumholz}
\affil{Research School of Astronomy and Astrophysics, Australian National University, Cotter Rd., Weston Creek
ACT 2611, Australia}
\affil{ARC Centre of Excellence for Astronomy in 3 Dimensions (ASTRO-3D), Canberra ACT 2601, Australia}

\author{Christoph Federrath}
\affil{Research School of Astronomy and Astrophysics, Australian National University, Cotter Rd., Weston Creek
ACT 2611, Australia}
\affil{ARC Centre of Excellence for Astronomy in 3 Dimensions (ASTRO-3D), Canberra ACT 2601, Australia}

\author{Mark Heyer}
\affil{Department of Astronomy, The University of Massachusetts, Amherst, MA, 01003, USA}

\author{Shivan Khullar}
\affil{Canadian Institute for Theoretical Astrophysics, 60 St. George Street, University of Toronto, Toronto ON M5S 3H4, Canada}
\affil{David A. Dunlap Department of Astronomy and Astrophysics, University of Toronto, 50 St George Street, Toronto, Ontario, M5S 3H4, Canada}

\author{S. Thomas Megeath}
\affil{Ritter Astrophysical Research Center, Department of Physics and Astronomy, University of Toledo, Toledo, OH
43606, USA}

\author{Philip C. Myers}
\affil{Center for Astrophysics, Harvard \& Smithsonian, 60 Garden Street, Cambridge, MA 02138, USA}

\author{Stella S. R. Offner}
\affil{Department of Astronomy, The University of Texas at Austin, 2500 Speedway, Austin, TX 78712, USA}

\author{Judith L. Pipher}
\affil{Department of Physics and Astronomy, University of Rochester, Rochester, NY 14627, USA}

\author{William J. Fischer}
\affil{Space Telescope Science Institute, Baltimore, MD 21218, USA}

\author{Thomas Henning}
\affil{Max-Planck-Institute for Astronomy, Königstuhl 17, D-69117 Heidelberg, Germany}

\author{Joseph L. Hora}
\affil{Center for Astrophysics, Harvard \& Smithsonian, 60 Garden Street, Cambridge, MA 02138, USA}



\begin{abstract}

One of the most important and well-established empirical results in astronomy is the Kennicutt-Schmidt relation between the density of interstellar gas and the rate at which that gas forms stars. A tight correlation between these quantities has long been measured at galactic scales. More recently, using surveys of YSOs, a KS relationship has been found within molecular clouds relating the surface density of star formation to the surface density of gas; however, the scaling of these laws varies significantly from cloud to cloud. In this Letter, we use a recently developed, high-accuracy catalog of young stellar objects from \textit{Spitzer} combined with high-dynamic-range gas column density maps of twelve nearby ($<$1.5 kpc) molecular clouds from \textit{Herschel} to re-examine the KS relation within individual molecular clouds. We find a tight, linear correlation between clouds' star formation rate per unit area and their gas surface density normalized by the gas free-fall time.
The measured intracloud KS relation, which relates star formation rate to the volume density, extends over more than two orders of magnitude within each cloud and is nearly identical in each of the twelve clouds, implying a constant star formation efficiency per free-fall time $\epsilon_{\rm ff}\approx 0.026$. The finding of a universal correlation within individual molecular clouds, including clouds that contain no massive stars or massive stellar feedback, favors models in which star formation is regulated by local processes such as turbulence or stellar feedback such as protostellar outflows, and disfavors models in which star formation is regulated only by galaxy properties or supernova feedback on galactic scales.

\end{abstract}


\section{Introduction} 
\label{sec:intro_sg}

In galactic disks, there is a well-established correlation between the gas mass and star formation rate per unit area, when both quantities are measured in kpc-scale or larger patches \citep[e.g.,][]{Kennicutt98, Bigiel08, Leroy13}; this correlation is known as the Kennicutt-Schmidt (KS) relation \citep{Schmidt59}. The correlation, however, worsens as one measures progressively smaller regions, and there is little correlation between the carbon monoxide and ionizing or far-infrared luminosities -- standard proxies for gas mass and star formation rate, respectively -- of individual molecular clouds or filaments $\lesssim 100$ pc in size \citep{Mooney88, Schruba10, Onodera10, Kruijssen14, Ochsendorf17, Zhang19}. This apparent lack of a correlation can be the result of a true spread in the star formation rate per unit mass among clouds \citep{Lee16}, or the failure of the proxies for mass and star formation rate. 
The latter is possible when we estimate star formation rate using the luminosity of massive stars, because this proxy may be under-sampled on small scales \citep{Calzetti12}, and also underestimates the true star formation rate until the stellar population is old enough ($\sim$5-10 Myr) to have reached a statistical steady state between the formation of new massive stars and the deaths of older ones \citep[e.g.,][]{Krumholz07}. 
Conversely, gas tracers also suffer from undersampling \citep{Calzetti12} and timescale issues: massive stars can rapidly disperse the gas from which they formed \citep[e.g.,][]{Chevance21a}, and if we observe a stellar population where dispersal is well underway, we will underestimate the mass of gas that was present when the stars formed. Thus analyses based on massive stars tend to underestimate the star formation rate per unit mass in young clouds and overestimate it in old clouds. When we measure the KS relation in kpc-scale patches, we average over large numbers of clouds at random ages, these errors cancel, and we recover the correct mean star formation rate per unit mass. However, the uncertainties for individual clouds might nonetheless be substantial, artificially creating scatter in the KS relation at smaller scales \citep{Feldmann11, Kruijssen14, Kreckel18a}.

Whether the observed large scatter in the KS relation at small scales indicates a real scatter in star formation rate per unit mass, or whether it is simply an artifact of the observational errors described above, has profound implications for our understanding of the mechanisms by which star formation is regulated. If it is real, this suggests that the KS relation on galactic scales is due to feedback processes acting at similar scales, most likely the balance between gravity and supernovae \citep{Ostriker11, Hopkins11, Faucher-Giguere13}, and individual clouds are free to collapse to stars with high efficiency; indeed, the lack of a KS relation within individual clouds is a direct prediction of such models \citep{Lee16}. On the other hand, if a KS relation does hold within single clouds, particularly those containing no stars massive enough to produce supernovae, this implies that some smaller-scale or more universal mechanism inhibits star formation within individual molecular clouds. These mechanisms include turbulence, magnetic fields \citep{Krumholz05, Federrath12} or stellar feedback in the form of protostellar outflows, stellar winds and ionizing radiation \citep{Krumholz12a, Federrath15, Duo20, Guszejnov21}.

A natural experiment for deciding between these possibilities is to search for a KS relation within individual molecular clouds using counts of the recently formed stars or protostars identified by their bright infrared emission from circumstellar dust. 
The rarity of massive stars and their disruptive effect on their host cloud means they are poor tracers on cloud scales. Protostars, by contrast, have the advantage that they sample a much shorter time interval and therefore provide a much better estimate of the ``instantaneous'' star formation rate (SFR), and they allow measurements of the SFR even in clouds that lack massive stars and have not been significantly affected by feedback. Studies based on this method generally do find a reasonable correlation between the number of young stellar objects (YSOs) in a cloud and its gas mass above a certain density, or its gas mass divided by its mean-density free-fall time \citep{Krumholz12, Lada12, Heyer16, Krumholz19}. Within molecular clouds, several studies have found a power-law correlation between the surface densities of YSOs and gas \citep{Gutermuth11, Lada13, Willis15}. Most recently, \cite{Pokhrel20}  used high accuracy YSO catalogs and high dynamic range gas column densities to show the presence of these laws in twelve nearby clouds. This correlation is consistent with a star formation surface density being proportional to the gas surface density squared. The scaling of this law, however, varies significantly between clouds.  Moreover, their analysis technique examines the density of gas around stars on a star-by-star basis, and therefore cannot easily determine whether there is a KS relation based on the volume density of gas. Using the same data, we apply a different approach to determine the star formation law that includes a dependence on the volume density of the gas. We find that the star formation law can be recast as an effectively universal linear dependence of the surface densities between star formation rate and gas mass per free-fall time, with a very less scatter between clouds.

\section{Observations} \label{sec:obs_sg}

The input data for our study consist of a matched set of protostellar catalogs and cloud column density maps. We use such matched catalogs and maps for the star-forming regions Ophiuchus, Perseus, Orion-A, Orion-B, Aquila-North, Aquila-South, NGC 2264, S140, AFGL 490, Cep OB3, Mon~R2, and Cygnus-X. For H$_{2}$ column density maps, we use \textit{Herschel}-derived column densities. For the clouds that are $<$500 pc distance, we used the column density maps from the $Herschel$ Gould Belt Survey \citep{Andre10}. Full details of the data reduction procedure for the clouds that are $>$500 pc away are provided in \cite{Pokhrel20}, but we summarize here for reader convenience. We construct the column density maps using \textit{Herschel/SPIRE} and \textit{Herschel/PACS} imaging at 160~$\mu$m, 250~$\mu$m, 350~$\mu$m, and 500~$\mu$m, convolved to a common resolution. In each pixel, we fit the observed spectrum using a model for dust emission in which the free parameters are the gas column density and the temperature; in these fits the dust opacity per unit mass at 500~$\mu$m is fixed to $\kappa_{500\,\mu\mathrm{m}} = 2.90$~cm$^{2}$~g$^{-1}$ based on the OH4 dust model of \citet{ossenkopf94}. Our column density maps are the results of these fits, and can be expressed equivalently in column of H$_2$ molecules, $N({\rm H}_2)$, or gas mass column $\Sigma_{\rm gas}$; the two are related by
\begin{equation}
    \Sigma_{\rm gas} = \frac{2 m_{\rm H}}{X} N({\rm H_2}),
\end{equation}
where $m_{\rm H} = 1.67\times 10^{-24}$~g is the mass of a hydrogen atom and $X = 0.71$ is the hydrogen mass fraction of the local interstellar medium \citep{Nieva12}. We also mask pixels where the estimated dust temperature exceeds a threshold value that indicates a Rayleigh-Jeans limit, since in this regime the column density estimate becomes very uncertain -- see \cite{Pokhrel20} for details. In the highest density regions the dust emission can be optically thick even at 500 $\mu$m and our estimation of column densities may represent the lower limits. However, we are not probing gas beyond N(H$_2$) $\sim$ 10$^{23}$ cm$^{-2}$, and the effect on our results is minimal. To the extent that optical depth effects are significant, they would cause us to slightly underestimate the gas mass at the highest column densities.

For protostars, we use the \textit{Spitzer} Extended Solar Neighborhood Archive (SESNA) catalog compiled by R.~Gutermuth et al.~(in preparation). SESNA is constructed using combined \textit{Spitzer} IRAC~\citep{Fazio2004} 3.6, 4.5, 5.8, 8.0~$\mu$m, MIPS \citep{Rieke2004} 24$\mu$m, and near-IR (1.24, 1.67, 2.16~$\mu$m) photometry from the Two Micron All-Sky Survey (2MASS; \citealt{Skrutskie2006}) spanning $\sim$90~deg$^2$. Near-IR photometry from the UK Infrared Deep Sky Survey Galactic Plane Survey (UKIDSS GPS \citealt{Lucas2008}) data was used exclusively for our most distant target, Cygnus-X. Sources with excess IR emission are distinguished from field stars and further subdivided into various YSO and contaminant classifications (e.g.,~background galaxies and unresolved molecular hydrogen shock emission) using a series of reddening-safe color and flux selections~\citep{Gutermuth09}. With a few exceptions \citep{Gutermuth11,Pokhrel20}, prior work on the intracloud KS relation employed protostar identifications that required 24~$\mu$m flux measurements (e.g., \citealt{Heiderman10, Evans14}). This requirement strongly limits protostar sensitivity due to confusion with resolved nebulosity and neighboring bright sources as are found in young stellar clusters \citep{Kryukova14,Megeath16,Gutermuth2015}.  SESNA and related Spitzer censuses of YSOs make robust protostar identifications that do not require 24 $\mu$m photometry, improving protostar completeness under these circumstances \citep{Gutermuth09,Megeath12}.  In addition, SESNA has a well-measured rate of contamination from extragalactic interlopers and edge-on disks \citep{Gutermuth08,Gutermuth09}, and we can therefore correct statistically for these contaminating effects. The correction procedure is explained in detail in \cite{Pokhrel20}; all our analysis in this work makes use of the statistically-corrected data.

\section{Methods} \label{sec:methods_sg}

\begin{figure}[ht]
    \centering
    \includegraphics[scale=0.35]{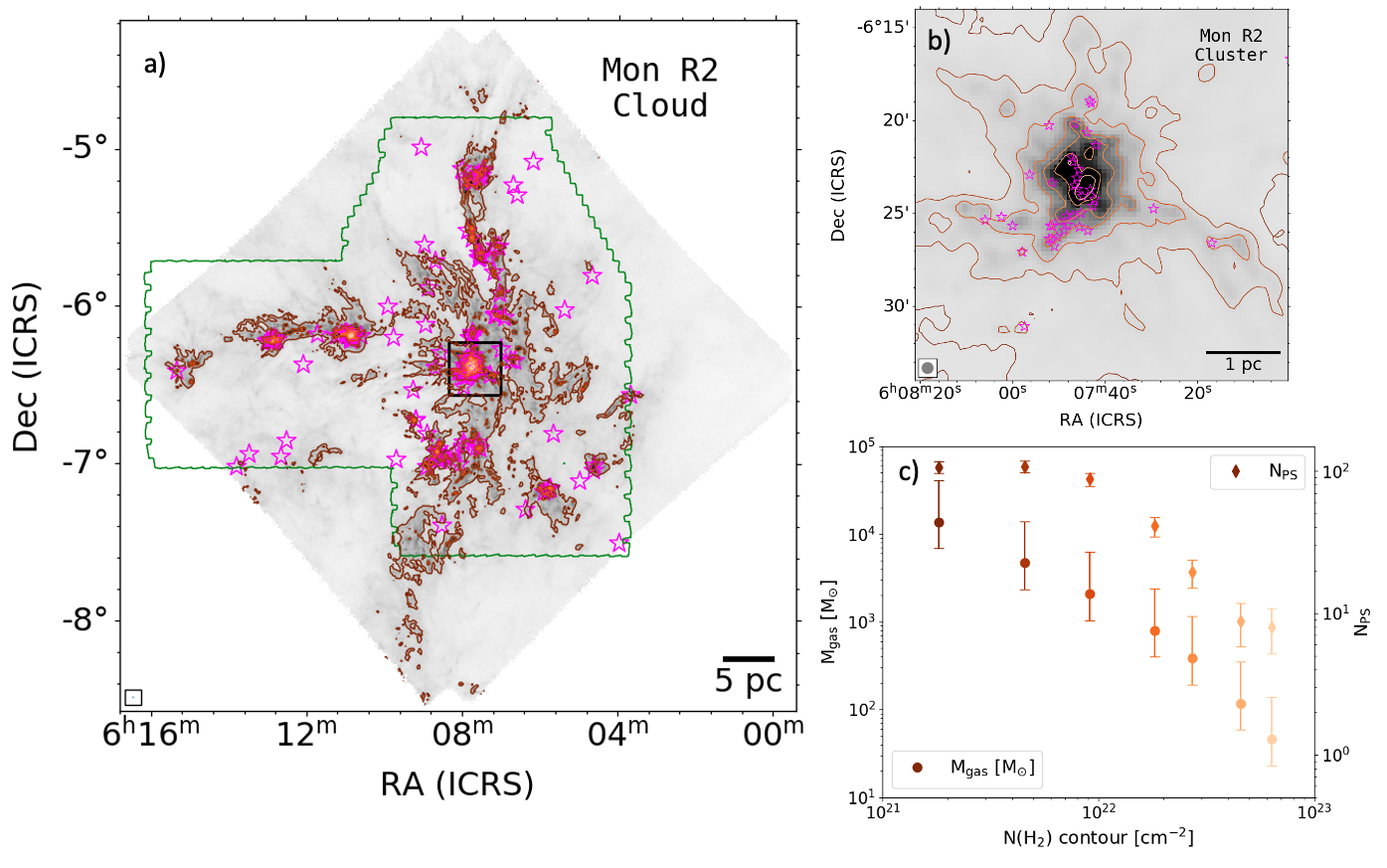}
    \caption{\textbf{a)}: Gas column density map of the Mon~R2 GMC derived from $Herschel$ observations \citep{Pokhrel16}. Green contours show the $Spitzer$ coverage map that is used for identifying protostars. The brown contours indicate molecular hydrogen column densities of $N({\rm H_2}) = (2, 5, 10, 20, 30, 50, 70) \times 10^{21}$ cm$^{-2}$, from lowest to highest. Protostars are shown as magenta stars. \textbf{b)}: Zoom-in view of the 5 $\times$ 5 parsec region centered at the Mon~R2 cluster that is shown as a black box in the left panel. \textbf{c)}: Gas mass and the number of protostars enclosed by each contour shown in panels (a) and (b). The colours of the points match the colours of the corresponding contours.}
    \label{fig:monr2}
\end{figure}

Given the input catalogs, we construct a series of contours within which we measure  the enclosed gas mass $M_{\rm gas}$, enclosed number of protostars $N_{\rm PS}$, and enclosed area $A$ (measured in physical rather than angular units). Our approach is similar to that explored by \citet{Heiderman10} and \citet{Lada10}. We place the lowest contour at the lowest value of $N({\rm H}_2)$ such that the resulting contour is entirely enclosed by the footprints of the SESNA catalog and the column density map. We then place additional contours with a uniform spacing corresponding to $0.5$ magnitudes in $A_V$, where for our OH4 dust model $0.5$~mag of extinction in V corresponds to a gas column $N({\rm H_2})\sim 5 \times 10^{20}$~cm$^{-2}$ until the smallest contour does not enclose any protostar. Our estimates for the minimum and the maximum A$_{\rm{V}}$ for each cloud are given in Table \ref{tab:fits}. The result of this procedure is a set of $(M_{\rm gas}, N_{\rm PS}, A)$ triples for each contour level in each cloud, which forms the basis for our analysis in this work. We show our data for one example cloud, Mon~R2 GMC, in Figure~\ref{fig:monr2}.

From our triple of directly measured quantities, we derive three additional quantities: the gas surface density $\siggas$, the star formation surface density $\sigsfr$, and the free-fall time $t_{\rm ff}$. The first of these is straightforward: $\siggas = M_{\rm gas} / A$. To derive $\sigsfr$, we adopt $M_{\rm PS} \approx 0.5\,M_\odot$ for the mean mass of protostars in our catalogue \citep{Evans09}, and the duration of the protostellar phase during which newborn stars will be included in our catalogue is $t_{\rm PS} \approx 0.5$~Myr \citep{,Dunham14,Dunham15}. Consequently, we compute the star formation rate within each contour as $\mbox{SFR} = N_{\rm PS} M_{\rm PS}/t_{\rm PS}$, and the star formation rate per unit area as $\sigsfr = \mbox{SFR}/A$. In order to estimate the free-fall time, we follow \citet{Krumholz12} in computing the density of the material within each surface density contour by assuming that the unseen dimension along the line of sight is comparable to the two dimensions observed in the plane of the sky, so that $\rho = 3\sqrt{\pi} M_{\rm gas}/4 A^{3/2}$; we then compute the free-fall time as $t_{\rm ff} = \sqrt{3\pi/32 G \rho}$. This amounts to assuming that the region being studied is a sphere in three dimensions.

For the best-fit analysis, we use the Orthogonal Distance Regression (ODR) method in \cite{Pokhrel20} as well as in all the best-fit analyses performed in this Letter. Hence, biases caused by different fitting techniques when comparing the results from the two studies are minimized. In the ODR method, uncertainties in both axes are used to find the regression line that is orthogonal to the residuals in finding optimized parameters. Thus, this method is preferred over the Ordinary Least Squares method for our analysis. For the details of using ODR in astronomical datasets, see \cite{Isobe90} and \cite{Akritas96}.

We estimate uncertainties on our derived quantities as follows. First, we find typical uncertainties of $\sim$30\% in the \textit{Herschel} derived column density maps and up to a factor of two uncertainty in the derived gas mass (see \citealt{Pokhrel16} for the details of uncertainty estimation). We propagated the uncertainties in column density to estimate uncertainties in derived $\siggas$. For the uncertainty in the number of protostars enclosed by each $N$(H$_2$) contour, we assume Poissonian errors so the error on $N_{\rm PS}$ is $\sqrt{N_{\rm PS}}$ \citep{Khullar19}, and propagate this to obtain the uncertainty in $\sigsfr$. Finally, \citet{Hu21} shows that the assumption of a uniform, spherical region that we use to estimate $t_{\rm ff}$ is likely responsible for adding a scatter of $\sim$0.2 dex. However, because this is a systematic rather than a random error, we do not attempt to propagate it below; we defer attempts to correct for this effect to Hu et al.~(2021b, in preparation).

\section{Results} \label{results}

\subsection{Variation of $\sigsfr$ with $\siggas$}

We begin by investigating the relationship between $\sigsfr$ and $\siggas$. In \citet{Pokhrel20}, we used the local YSO density at the location of protostars, as given by an $n^{\rm{th}}$ nearest neighbor density to measure $\sigsfr$ and the gas column density at that location to determine $\siggas$. We found that for each of our 12 molecular clouds, $\sigsfr \propto \siggas^2$. Thus, the analysis in \citet{Pokhrel20} is different from the one we perform here, in that \citet{Pokhrel20} examine the gas surface density around each protostar (i.e., a star-centric analysis), whereas here we are investigating the properties of regions defined by the clouds column density; the latter has the advantage that it allows us to investigate the dependence of the intracloud KS relation on cloud volume density.

In Figure~\ref{fig:stargas}a, we plot the relationship between $\Sigma_{\rm SFR}$ and $\Sigma_{\rm gas}$ as defined by our contours. Clearly, the relation is approximately linear in log-log, and we report the best-fit results of the data to a linear functional form in Table~\ref{tab:fits}. We fit only the data that comes from column density contours $<3\times 10^{22}$~cm$^{-2}$, since above this limit the contours and number of protostars enclosed become very small, and Poisson errors in $\Sigma_{\rm SFR}$ become large. Considering all 12 clouds, the average best-fit slope is $2.00 \pm 0.27$ and the average best-fit y-intercept is $-4.11 \pm 0.80$. Near the center of the observed data range at a gas surface density $\Sigma_{\rm gas}=10^{2.5}$~$M_\odot$~pc$^{-2}$, the standard deviation of the measured values of $\log\Sigma_{\rm SFR}$ across all clouds is $0.30$. We show the line corresponding to our average best-fit parameters, with this scatter, in Figure~\ref{fig:stargas}a. Individual scatter that is intrinsic to an individual cloud is not considered when calculating the standard deviation as they may be caused by observational uncertainties, while cloud-to-cloud scatter is more robust. For the star-centric approach in \cite{Pokhrel20}, we used the best-fit equations for each cloud (c.f. Table 3 in \citealt{Pokhrel20}) and find the standard deviation of measured $\log\Sigma_{\rm SFR}$ to be $0.33$. Furthermore, the average best-fit slope in \cite{Pokhrel20} is $2.02 \pm 0.20$ and the average best-fit y-intercept is $-3.88 \pm 0.59$. The best-fit results in these two approaches are well within 1-$\sigma$ standard deviation.
Such stark similarities in results using two distinct methods is strong evidence that correlation is not being biased by the method.

\subsection{Variation of $\sigsfr$ with $\siggas/t_{\rm{ff}}$}

While the surface densities of gas and star formation are the quantities most directly accessible from observations, most theoretical models that predict the existence of a KS relation for single clouds predict a dependence on the gas free-fall time \citep{Krumholz05,Padoan12,Federrath13,Krumholz19}, which depends on the volume density. Incorporating the free-fall time also gives a tighter correlation when measuring the cloud-to-cloud KS relation \citep{Krumholz12, Heyer16}. Inclusion of the volume density and normalization of $\siggas$ by the free-fall time (t$_{\rm{ff}}$) is the primary difference of our approach over \cite{Pokhrel20}.

\begin{figure}[htbp]
\centering
\includegraphics[scale=0.38]{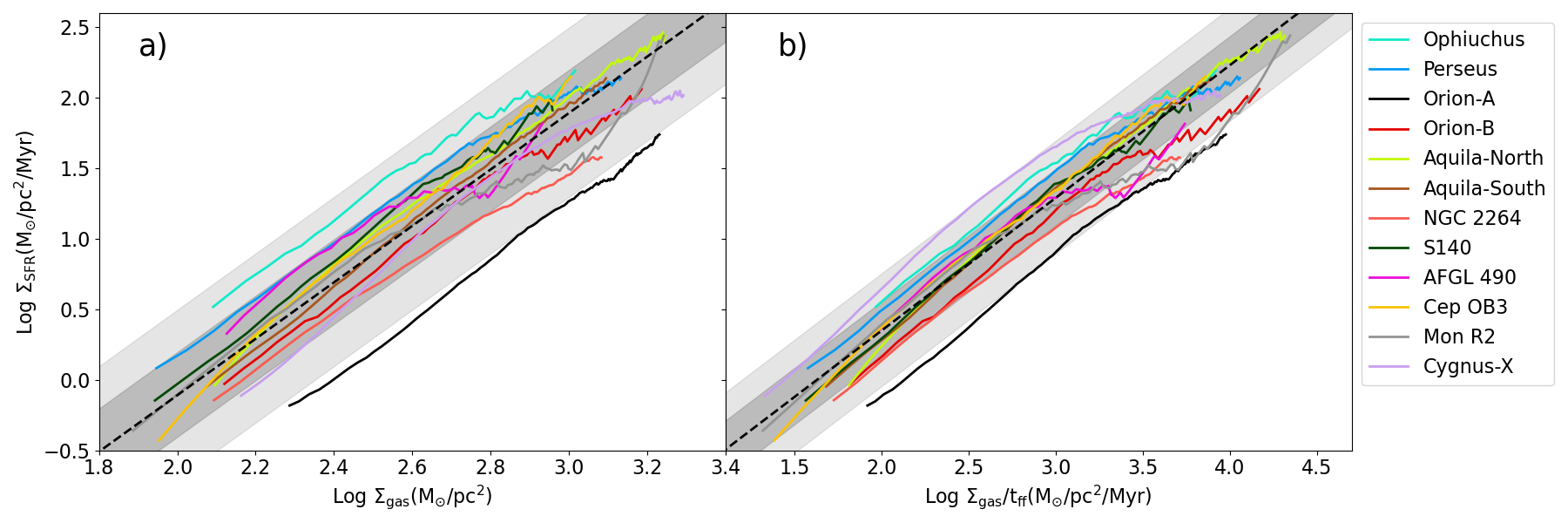}
\caption{a) $\log \Sigma_{\rm{SFR}}$ vs.~$\log \Sigma_{\rm{gas}}$ for contours defined on each of the 12~sample clouds (as indicated in the legend) b) Same as (a), but using $\Sigma_{\rm{gas}}/t_{\rm{ff}}$ on the horizontal axis. In both panels, black dashed lines show the median best fit relation, using the parameters shown in Table \ref{tab:fits}; for (b), the black dashed line shows the fit constrained to have a slope of unity, though the best fit for an unconstrained slope is nearly indistinguishable. The darker shaded region shows the standard deviation of the data (see Table~\ref{tab:fits}) around the average best fit line. and the lighter shaded region represents two times the standard deviation.
}
\label{fig:stargas}
\end{figure}

Figure~\ref{fig:stargas}b shows the relationship between $\Sigma_{\rm SFR}$ and $\Sigma_{\rm gas}/t_{\rm ff}$, and Table~\ref{tab:fits} shows the results of fitting a power law relationship between these quantities. It is immediately clear that the scatter is much smaller for this relationship than for the one between $\Sigma_{\rm SFR}$ and $\Sigma_{\rm gas}$ alone; quantitatively, the standard deviation of the $\Sigma_{\rm SFR}$ is reduced from $0.30$ to $0.21$ (computed at $\Sigma_{\rm gas}/t_{\rm ff} = 10^{2.5}$~$M\odot$~pc$^{-2}$~Myr$^{-1}$) by inclusion of the free-fall time. Moreover, the relationship is now linear, with a median best-fit slope of $0.99$. This finding, coupled with the theoretical predictions for a linear relationship, motivates us to carry out a fit where we fix the slope to unity and fit only the offset, so the functional form is
\begin{equation}
    \log \Sigma_{\rm SFR} = \log\left(\Sigma_{\rm gas}/t_{\rm ff}\right) + \log \epsilon_{\rm ff},
\end{equation}
where $\epsilon_{\rm ff}$ is the fraction of the gas mass converted to stars per free-fall time. The resulting fits are indistinguishable within the error bars from those where we allow the slope to vary (see Table~\ref{tab:fits}). We show the fit using the median value of $\epsilon_{\rm ff} \approx 0.026$ in Figure~\ref{fig:stargas}b.

The reformulation of the intracloud KS relation in terms of $\log \sigsfr$ and $\log \siggas/t_{\rm{ff}}$ has the advantage of both a linear dependence between the quantities and a lower dispersion between clouds. The linear correlation between $\sigsfr$ and $\siggas/t_{\rm{ff}}$ implies that we can reformulate this as a relationship between the volume density of star formation and volume density of gas. Since the volume density is the more fundamental physical quantity for determining fragmentation scales and collapse times, and since the scatter between clouds is comparatively low, we propose that this KS law is a more fundamental, universal formulation of the intracloud KS relation.

\subsection{Variation of $\epsilon_{\rm ff}$ with $\siggas$} \label{epsff}

\begin{figure}[ht]
    \centering
    \includegraphics[scale=0.45]{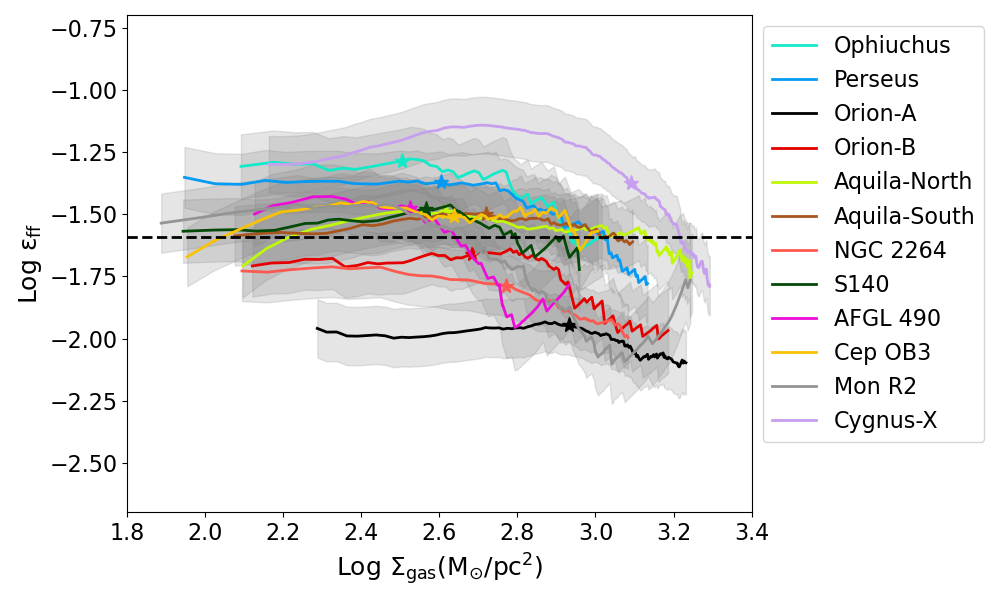}
    \caption{Variation of the free-fall efficiency ($\epsilon_{\rm{ff}}$) with $\Sigma_{\rm{gas}}$ for our sample of clouds. The shaded regions are the uncertainties in $\epsilon_{\rm{ff}}$ which are computed by assuming a Poisson distribution for the number of protostars \citep{Khullar19}. The stars along the curves for each cloud indicate the column density for which t$_{\rm ff}$ $\approx 0.5$ Myr. The median of the logarithm of $\epsilon_{\rm{ff}}$ ($-1.59$, see Table~\ref{tab:fits}) is shown by a black dashed line.}
    \label{fig:freefall}
\end{figure}

To demonstrate that the correlations seen in Figure \ref{fig:stargas}b are not just an artifact created by comparing two quantities that are inversely proportional to the area, we further examine the value of $\epsilon_{\rm ff}$ as a function of column density for each of our clouds in Figure~\ref{fig:freefall}. We construct this figure following the method of \citet{Khullar19}, whereby we vary the contour level as shown in Figure~\ref{fig:monr2}, and within each contour we measure $\epsilon_{\rm ff} = \textrm{SFR}/(M_{\rm gas}/t_{\rm ff})$, where the values of $\textrm{SFR}$, $M_{\rm gas}$, and $t_{\rm ff}$ are the values within the contour. The Figure shows how $\epsilon_{\rm ff}$ varies with mean gas column density within the corresponding contour $\Sigma_{\rm gas}$.

Note that $t_{\rm ff}\propto A^{3/4}$, while $\siggas \propto A^{-1}$, so if the correlation shown in Figure~\ref{fig:stargas}b were primarily due to the fact that both axes depend similarly on area, then in Figure~\ref{fig:freefall} we would expect to find $\epsilon_{\rm ff} \propto \siggas^{-3/4}$. Figure~\ref{fig:freefall} clearly shows no such correlation, which strongly indicates that the correlation shown in Figure~\ref{fig:stargas}b is real rather than spurious. 
We also find no evidence for any threshold at which star formation becomes efficient, i.e., where $\epsilon_{\rm ff}$ rises substantially. This is contrary to some earlier analyses using much more limited data \citep{Lada10,Heiderman10,Konyves15a}. Instead we find that in almost all clouds $\epsilon_{\rm ff}$ is nearly constant over $\approx 1$ decade in column density from $\approx 100$--$1000$~$M_\odot$~pc$^{-2}$, and that at column densities $\gtrsim 1000$~$M_\odot$~pc$^{-2}$ the value of $\epsilon_{\rm ff}$ decreases rather than increases.

The decrease in $\epsilon_{\rm ff}$ at high column density is contrary to the naive expectation that star formation should become more rather than less efficient in denser gas. However, it seems likely that the drop in apparent $\epsilon_{\rm ff}$ is not indicative of a true decline in star formation efficiency but is rather a result of one of three possible effects. One is that the YSOs we use to estimate the star formation rate and thence $\epsilon_{\rm ff}$ average over a finite timescale of $t_{\rm YSO}\approx 0.5$ Myr, and this can induce bias in estimates of $\epsilon_{\rm ff}$ at high density. We discuss this in more detail in \autoref{ssec:biases}.

A second possible explanation is that protostellar lifetimes might not be independent of density as we have assumed. Protostellar luminosities are observed to be higher in dense regions of molecular clouds \citep{Kryukova14,Dunham14, Kirk17}, and it is possible that this is a signature of more rapid accretion that could, in turn, lead to more rapid progression through the evolutionary phase selected from our source catalog. In this case, our method would lead us to somewhat underestimate $\epsilon_{\rm ff}$ in the densest regions we survey.

A third possible explanation is that the densest parts of star-forming regions are also sites of bright and complex emission in the infrared that can contribute to locally reduced YSO sensitivity \citep{Megeath16}. Our SENSA catalog is more sensitive than previous ones in these regions, as we discuss in the next section, but we may still suffer from some incompleteness in the densest regions. Again, this would cause us to underestimate $\epsilon_{\rm ff}$ in those regions, a feature that is consistently observed across all clouds in this analysis.

\section{Discussion}

\subsection{Evolutionary biases in $\epsilon_{\rm ff}$}
\label{ssec:biases}

As discussed in \autoref{sec:intro_sg}, the primary motivation for this study is to circumvent the biases inherent in studying star formation using tracers based on massive stars, which integrate over relatively long periods of $5-10$ Myr, and likely alter the star-forming environment over such timescales. It is therefore important to investigate to what extent our results may suffer from similar evolutionary biases. Feedback effects from our low-mass protostars are likely small, but counting protostars still amounts to measuring the star formation rate integrated over a finite time $t_{\rm YSO}\approx 0.5$ Myr. If either gas or stellar quantities evolve on this timescale, this could cause an error in our estimates of $\epsilon_{\rm ff}$. For example, if the gas were collapsing such that the density has increased over the past $\sim 0.5$ Myr, then the present-day density that we measure is higher than the mean density at the time when the YSOs formed, in which case we are underestimating $t_{\rm{ff}}$ and thus overestimating $\epsilon_{\rm{ff}}$. Similarly, if YSOs born inside one of our contours were to move out of it during our $\sim 0.5$ Myr integration interval, then we would underestimate the SFR and thus $\epsilon_{\rm ff}$.

The regions where we expect evolutionary effects to be significant correspond to those for which the free-fall time, $t_{\rm ff}$, is comparable to the integration time, $t_{\rm YSO}\approx 0.5$ Myr. This is because the free-fall time is both the fastest timescale over which gas properties are likely to change (e.g., becoming denser due to collapse), and the fastest timescale over which we expect YSO motion to be significant.\footnote{The reason that YSO motion is related to the free-fall time is that the natural timescale for YSO motion is the crossing time, and for a region with virial parameter $\alpha_{\rm vir}\sim 1$, this is roughly equal to the free-fall time \citep[e.g.,][]{Krumholz07}.} Thus evolutionary biases are a potential concern wherever $t_{\rm ff} \lesssim t_{\rm YSO}$. In \autoref{fig:freefall}, we mark the $\Sigma_{\rm gas}$ contour at which $t_{\rm{ff}}$ $\sim$ 0.5 Myr with a star. We see that, for most clouds, the decline in $\epsilon_{\rm{ff}}$ at higher $\Sigma_{\rm{gas}}$ begins close to the marked point, which is strongly suggestive that evolutionary effects may be the reason that we see the decline in $\epsilon_{\rm{ff}}$ at higher $\Sigma_{\rm{gas}}$, in addition to the two observational reasons (non-constant protostellar lifetimes and incompleteness) discussed in \autoref{epsff}. However, the converse conclusion also applies: evolution should not be a concern for lower $\Sigma_{\rm{gas}}$ regions where $t_{\rm{ff}}\gtrsim 0.5$ Myr. Even if we limit ourselves to the parts of the $\epsilon_{\rm ff}$ curves that lie to the left of the stars in \autoref{fig:freefall}, we still find that $\epsilon_{\rm ff}$ both varies little from cloud-to-cloud, and is nearly constant over almost an order of magnitude dynamic range in $\Sigma_{\rm gas}$.

\subsection{Comparison with previous cloud-scale studies} \label{app:comp}

\begin{figure}[ht]
    \centering
    \includegraphics[scale=0.45]{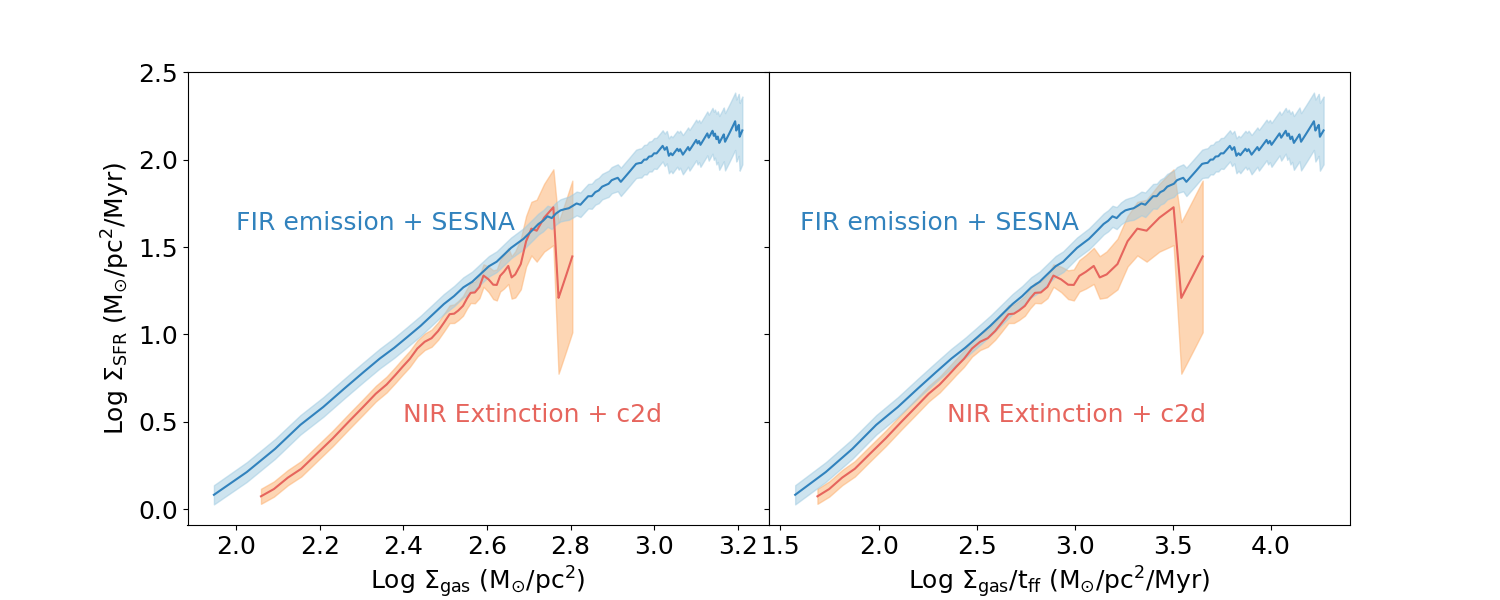}
    \caption{Comparison of star--gas surface density correlation plots between our data (far-IR $Herschel$ $N({\rm H}_2)$ gas and SESNA protostars; shown in blue) and a mid-IR extinction map and together with protostars from the c2d catalog (shown in red) for the Perseus molecular cloud.}
    \label{fig:c2dsesna}
\end{figure}

As discussed in the introduction, we are not the first authors to search for a single-cloud KS relation (see \citealt{Gutermuth11, Lada13, Willis15} and so on). Using the c2d and Gould Belt $Spitzer$ Legacy Program, 
\cite{Evans14} found a correlation between $\sigsfr$ and $\Sigma_{\rm gas}/t_{\rm ff}$, but with more scatter and a steeper slope of $\sim$1.47 for an ensemble of different clouds. It is therefore of interest to understand why we find a much distinct intracloud KS relation in \cite{Pokhrel20} and in this Letter. The primary explanation for the difference is the sensitivity and depth of our \textit{Herschel}-derived column density maps, with the depth of the SENSA YSO catalog as a secondary factor. To demonstrate this, we focus on the Perseus cloud as an example, and repeat our analysis using a column density map derived from extinction together with the c2d protostellar catalog (both from \citealt{Evans09}); these are representative of the data quality available in earlier studies. We plot the correlation between $\Sigma_{\rm SFR}$, $\Sigma_{\rm gas}$, and $\Sigma_{\rm gas}/t_{\rm ff}$ derived from these data in Figure~\ref{fig:c2dsesna}; the Figure also shows our results derived from \textit{Herschel} plus SESNA for comparison. The most obvious difference is that the older data cover a much smaller dynamic range -- $\lesssim 0.5$ decades in $\Sigma_{\rm gas}$, and $\lesssim 1.5$ decades in $\Sigma_{\rm gas}/t_{\rm ff}$, compared to $\gtrsim 1$ decade in $\Sigma_{\rm gas}$ and $\gtrsim 2$ decades in $\Sigma_{\rm gas}/t_{\rm ff}$ for our data. The difference is primarily a result of the extinction maps saturating at high column density, which prevents them from measuring the high values of $\Sigma_{\rm gas}$ that we can probe using far-infrared dust emission \citep{Pokhrel16}. A secondary contributor is that the SESNA YSO catalog is more complete in high-density regions. Furthermore, we have included a larger number of both low-mass star-forming clouds and high-mass star-forming GMCs in our sample and fit each cloud separately.
In contrast, \cite{Evans14} combined measurements from multiple clouds into a single fit and small differences between clouds may have affected the slope. For these reasons, the linear KS relation between $\Sigma_{\rm SFR}$ and $\Sigma_{\rm gas}/t_{\rm ff}$ apparent in our data was not favored by their analysis.

\subsection{Comparison with previous Galactic-scale and extragalactic studies}

In addition to comparing to previous searches for single-cloud KS relations, it is helpful to put our study in the context of whole-Galaxy and extragalactic studies of the star formation relation. We cannot meaningfully compare to studies using the traditional KS relation expressed in terms of surface densities of gas and star formation, since the surface density measured for a single, spatially-resolved, beam-filling molecular cloud, as in our data, is not the same as the surface density measured in a beam containing many clouds with a filling factor $\ll 1$, the typical situation in extragalactic studies. However, we can meaningfully compare distributions of $\epsilon_{\rm ff}$, since for a collection of equal-density clouds partly filling an observed beam, the measured value of $\epsilon_{\rm ff}$ for the whole beam is simply the SFR-weighted mean $\epsilon_{\rm ff}$ of the individual clouds. As discussed in \autoref{sec:intro_sg}, distributions of $\epsilon_{\rm ff}$ measured in Galactic-scale and extragalactic studies depend systematically on the size scale. Observations that average over regions of a few hundred pc or larger generally yield fairly small dispersions of star formation efficiency; for example \citet{Leroy17} and \citet{Utomo18} find $\sigma_{\log\epsilon_{\rm eff}} = 0.2 - 0.3$ dex for galaxies in the PHANGS sample. By contrast, measurements on $\sim 10-100$ pc scales yield contradictory results, with some reporting much larger dispersions than on larger scales; for example \citet{Lee16} find $\sigma_{\log\epsilon_{\rm ff}}\approx 0.8-0.9$ dex for individual molecular clouds in the Milky Way, while \citet{Ochsendorf17} find obtain $\sigma_{\log\epsilon_{\rm ff}}\approx 0.6$ dex in the LMC. Others, for example \citet{Vutisalchavakul2016} and \citet{Barnes2017}, report smaller dispersions that are closer to those found on larger scales. Our measured dispersion, $\sigma_{\log\epsilon_{\rm eff}} = 0.18$ dex on $\sim 1 - 10$ pc scales, is firmly in the small dispersion camp, and it is interesting to ask why.

One potential explanation might be that the clouds in our study are unrepresentative of those sampled in the extragalactic or Galaxy-scale studies, which are dominated by massive star-forming regions. However, we can quickly rule out this possibility. Our sample does include a number of massive star-forming regions (Orion~A, Cep~OB3, Mon~R2, Cygnus-X) that would be readily visible to Galaxy-scale or extragalactic studies; indeed, Cygnus-X contains $>10^6$ $M_\odot$ of molecular gas and $>2\times 10^4$ YSOs, which would place it the top quartile of \citeauthor{Lee16}'s Milky Way GMC catalog or \citeauthor{Ochsendorf17}'s LMC catalog by both mass and star formation rate. Moreover, both the GMC mass function \citep[e.g.,][]{Williams97, Heyer16} and the star cluster mass function \citep[e.g.,][]{Whitmore14} are relatively flat, $dN/dM\sim M^{-1.7}$ and $M^{-2}$, respectively. This implies that small clouds and star clusters make a non-negligible contribution to the integrated total gas mass or star formation rate measured in partly-filled beams; quantitatively, for a GMC mass function with slope $-1.7$ and a mass range of $10^3 - 10^6$ $M_\odot$, roughly 25\% of the mass is contained in clouds below $10^{4.5}$ $M_\odot$, which are the majority of our sample. Thus we cannot attribute the difference in $\sigma_{\log \epsilon_{\rm ff}}$ between our study and earlier Galactic-scale or extragalactic cloud studies to differences in the clouds being sampled.

Instead, a more likely explanation is that the large dispersion reported in earlier cloud-scale studies is simply an artifact of the observational errors inherent in measuring star formation rates using massive star-formation tracers, which integrate over relatively long timescales and thus sample times after which the star-forming environment has been significantly transformed by feedback. Indeed, there are already hints toward such a conclusion in the literature. For example, \citet{Gutermuth09, Gutermuth11} show that, for individual clouds, infrared luminosity does not correlate to the number of YSOs to better than an order of magnitude. \citet{Heyer16} measure $\epsilon_{\rm ff}$ in ATLASGAL clumps using intermediate-mass YSO counts rather than ionizing or infrared luminosity, and find $\sigma_{\log\epsilon_{\rm ff}}\approx 0.4$ dex, closer to both our results and the large-scale extragalactic results than to the cloud-scale measurements using ionizing or infrared luminosity. Moreover, results based on the latter two tracers appear to depend sensitively on exactly how one assigns SFRs to individual clouds: for example, the difference in dispersion for Milky Way GMCs reported by \citet{Vutisalchavakul2016} compared to \citet{Lee16} is almost entirely due to such differences, and in \citet{Ochsendorf17}'s study of the LMC, simply using counts of massive YSOs ($\gtrsim 8$ $M_\odot$) rather than H$\alpha$ luminosity as a star formation rate indicator, while leaving all other aspects of the analysis unchanged, reduces $\sigma_{\log \epsilon_{\rm ff}}$ by $0.1-0.2$ dex \citep{Krumholz19}. Together with our results here, these studies support the hypothesis that the primary explanation for the large scatter reported in some previous cloud-scale estimates of $\epsilon_{\rm ff}$ is a failure of ionizing and IR luminosity as a tracer, rather than a physical change in the star formation process in going from galactic to cloud scales; instead, the same mechanisms regulate star formation at size scales from $\sim 1 - 1000$ pc. This hypothesis will be directly testable in the next few years using \textit{JWST}, which will be able to detect YSOs at substantially larger distances than \textit{Spitzer}. If our hypothesis based on this study is correct, then repeating earlier cloud-scale studies using \textit{JWST}-detected YSOs rather than ionizing or IR luminoisty as star formation rate indicators should yield substantially lower dispersions in $\epsilon_{\rm ff}$.

\subsection{On the uniformity of $\epsilon_{\rm{ff}}$}

Our sample consists of clouds whose masses and SFRs span multiple orders of magnitude, yet we find that all clouds have roughly constant $\epsilon_{\rm{ff}}$, both from cloud to cloud and within a single cloud. This strongly suggests that star formation is regulated by local processes that are present in both low-mass and high-mass star forming regions. One candidate is magnetized, supersonic turbulence stirred and aided by feedback from low-mass stars \citep[e.g.,][]{Krumholz05, Krumholz12, Padoan12, Federrath12}. Even in low-mass star-forming regions, outflows can drive and maintain turbulence at parsec scales \citep{Bally16,Offner17}, and modern simulations including turbulence, magnetic fields, protostellar outflows, and thermal radiation from low-mass stars -- all processes that would be present even in our low-mass clouds -- yield $\epsilon_{\rm ff}$ values of a few percent, roughly consistent with our measurements \citep{Federrath15, Cunningham18, Li18}. This is a plausible explanation for our findings. In this view, it is also possible that the bend in $\epsilon_{\rm ff}$ we see at the highest surface densities is associated with the transition from supersonic to subsonic turbulence (e.g., \citealt{Federrath21}), since observations suggest that the surface densities at which we see the bend correspond roughly to those where the role of thermal motion in supporting the clouds begins to increase (e.g., \citealt{Pokhrel18}).

\section{Conclusions} \label{conclusion}

We use $Herschel$-derived H$_2$ column density maps and the SESNA YSO catalog to explore the intracloud KS relation in star-forming molecular clouds that are $<$1.5 kpc away. Our main conclusions are summarized below.

\begin{enumerate}
    \item We find that $\sigsfr~\propto~\siggas^2$ in all the clouds in our sample. The result is consistent with that reported by \cite{Pokhrel20}, who use a different and complementary analysis technique.
    \item Incorporating volume density reduces the scatter between different clouds and reveals a linear relation: $\sigsfr$ = $\epsilon_{\rm{ff}} \siggas/t_{\rm{ff}}$, where the proportionality constant $\epsilon_{\rm{ff}}$ is the free-fall efficiency.
    \item $\epsilon_{\rm{ff}}$ stays nearly constant and is independent of $\siggas$ in all the clouds. We find a median $\epsilon_{\rm ff} \approx 0.026$ and the cloud-to-cloud standard deviation of $\log \epsilon_{\rm{ff}}$ is $\approx 0.18$.
\end{enumerate}

Our results demonstrate that star formation within individual molecular clouds follows a tight KS relation at parsec scales, characterized by a linear relationship between star formation rate and mass normalized by free-fall time. 
This relationship is essentially the same in all the molecular clouds we studied. 
This is significant because the clouds themselves span a huge range of properties: for example, the Perseus and Ophiuchus clouds contain no stars with significant ionizing luminosities or winds, Cep OB3 and Mon R2 are sites of ongoing massive star formation, and Cygnus-X is comparable to large complexes observed in other galaxies. The latter two are comparable to the star-forming regions that are probed in extragalactic observations (at least for very nearby galaxies), while the former would be below the detection threshold of extragalactic star-formation studies.

The small scatter in log~$\epsilon_{\rm ff}$ we have measured rules out models in which star formation is regulated only at galactic scales, and not within individual clouds. For example, \cite{Murray15} propose that molecular clouds are collapsing and that, as a result, the star formation rate within them increases with time as $\mbox{SFR}\propto t^2$; \cite{Lee16} show that the observed dispersion in $\log\epsilon_{\rm ff}$ predicted by this model is $0.54$, a factor of $\approx 3$ larger than we observe. By contrast, a model in which there is no collapse and thus $\epsilon_{\rm ff}$ does not increase yields a dispersion in $\log \epsilon_{\rm ff}$ of 0.16, very close to what we observe. Thus our observations strongly favor the existence of a mechanism that keeps $\epsilon_{\rm ff}$ close to constant across all local molecular clouds. Moreover, this mechanism must not depend on the feedback provided by massive stars such as radiation and winds, since many of the clouds we have observed contain no massive stars.

\acknowledgments

We gratefully acknowledge funding support for this work from NASA ADAP awards NNX11AD14G (R.A.G.), NNX13AF08G (R.A.G.), NNX15AF05G (R.A.G., R.P.), 80NSSC18K1564 (R.P., S.T.M.), and NNX17AF24G (R.A.G., R.P.), and from the Australian Research Council awards FT180100375 (M.R.K.), DP190101258 (M.R.K.), DP170100603 (C.F.), FT180100495 (C.F.), and CE170100013 (M.R.K., C.F.). S.S.R.O. acknowledges NSF CAREER grant 1748571. This work is based in part on observations made with the $Spitzer$ Space Telescope, which was operated by the Jet Propulsion Laboratory, California Institute of Technology under a contract with NASA. Support for this work was provided by NASA through an award issued by JPL/Caltech. This work also uses observations made with $Herschel$, a European Space Agency cornerstone mission with science instruments provided by European-led Principal Investigator consortia and with significant participation by NASA.
\vspace{5mm}

\facilities{$Herschel$ (SPIRE and PACS), 2MASS, $Spitzer$ (IRAC and MIPS)}

\software{APLpy \citep{Robitaille12}, astropy \citep{Astropy13}, Matplotlib \citep{Hunter07}, NumPy (https://doi.
org/10.1109/MCSE.2011.37),
SciPy \citep{Jones01}.
          }

\clearpage
\begin{table}[ht]
\movetabledown=1.8in
\begin{rotatetable}
\centering
\begin{tabular}{|l|cc|cc|cc|c|}
\hline\hline
\multirow{2}{*}{Cloud}
& \multicolumn{2}{|c|}{A$_{\rm V}$ level (mag)}
& \multicolumn{2}{|c|}{$\log\Sigma_{\rm SFR} = a\log\Sigma_{\rm gas} + b$}
& \multicolumn{2}{c|}{$\log\Sigma_{\rm SFR}=a\log\left(\Sigma_{\rm gas}/t_{\rm ff}\right)+b$}
& $\log\Sigma_{\rm SFR}=\log\left(\Sigma_{\rm gas}/t_{\rm ff}\right) + \log\epsilon_{\rm ff}$
\\
& Min & Max & $a$ & $b$ & $a$ & $b$ & $\log \epsilon_{\rm ff}$ \\ \hline
Ophiuchus & 3.0 & 55.5 & 1.83$\pm$0.03 & -3.24$\pm$0.08 & 0.87$\pm$0.02 & -1.11$\pm$0.05 & -1.44$\pm$0.12 \\ 
Perseus & 2.0 & 92.0 & 1.88$\pm$0.02 & -3.54$\pm$0.05 & 0.92$\pm$0.01 & -1.31$\pm$0.03 & -1.51$\pm$0.13 \\ 
Orion-A  & 3.0 & 99.5 & 2.14$\pm$0.01 & -5.13$\pm$0.03 & 1.04$\pm$0.01 & -2.20$\pm$0.01 & -2.00$\pm$0.04 \\ 
Orion-B  & 3.0 & 90.5 & 2.14$\pm$0.01 & -4.58$\pm$0.04 & 1.01$\pm$0.01 & -1.84$\pm$0.02 & -1.77$\pm$0.13 \\ 
Aquila-North  & 4.0 & 99.5 & 2.08$\pm$0.03 & -4.23$\pm$0.07 & 1.01$\pm$0.01 & -1.71$\pm$0.04 & -1.60$\pm$0.07 \\ 
Aquila-South  & 3.0 & 99.5 & 2.20$\pm$0.01 & -4.62$\pm$0.02 & 1.03$\pm$0.01 & -1.75$\pm$0.02 & -1.55$\pm$0.03 \\ 
NGC 2264  & 3.0 & 99.5 & 1.71$\pm$0.02 & -3.64$\pm$0.05 & 0.86$\pm$0.01 & -1.55$\pm$0.03 & -1.85$\pm$0.09 \\ 
S140  & 2.0 & 53.0 & 2.09$\pm$0.02 & -4.17$\pm$0.06 & 0.97$\pm$0.01 & -1.61$\pm$0.03 & -1.57$\pm$0.06 \\ 
AFGL 490  & 4.0 & 40.0 & 1.44$\pm$0.06 & -2.55$\pm$0.15 & 0.67$\pm$0.03 & -0.82$\pm$0.08 & -1.70$\pm$0.21 \\ 
Cep OB3  & 3.0 & 57.0 & 2.36$\pm$0.02 & -4.93$\pm$0.05 & 1.03$\pm$0.01 & -1.73$\pm$0.03 & -1.52$\pm$0.04 \\ 
Mon~R2  & 2.0 & 94.0 & 1.71$\pm$0.04 & -3.41$\pm$0.09 & 0.82$\pm$0.02 & -1.25$\pm$0.05 & -1.80$\pm$0.24 \\ 
Cygnus-X  & 3.0 & 99.5 & 2.38$\pm$0.02 & -5.25$\pm$0.06 & 1.06$\pm$0.01 & -1.49$\pm$0.02 & -1.37$\pm$0.19 \\ 
\hline
Median  & & & 2.08 & -4.20 & 0.99 & -1.58 & -1.59 \\ 
Mean  & & & 2.00 & -4.11 & 0.94 & -1.53 & -1.64 \\ \hline
Spread  & & & \multicolumn{2}{|c|}{$\sigma_{\log\Sigma_{\rm SFR}} = 0.30$} & \multicolumn{2}{|c|}{$\sigma_{\log\Sigma_{\rm SFR}} = 0.21$} & $\sigma_{\log\epsilon_{\rm ff}}$ = 0.18 \\
\hline\hline
\end{tabular}
\caption{\label{tab:fits}
Best-fit parameters with errors. The column A$_{\rm V}$ indicates the range in visual extinction over which the relationship is measured. In each remaining group of columns, the equation at the top indicates the functional form being fit, and the columns under it give the best-fit parameters for each cloud; numerical values given are for $\Sigma_{\rm gas}$ in units of $M_\odot$~pc$^{-2}$, $t_{\rm ff}$ in units of Myr, and $\Sigma_{\rm SFR}$ in units of $M_\odot$~pc$^{-2}$~Myr$^{-1}$. 
The entries at the bottom of the table give the median and mean of the fit values, and the spread in the data as measured by their standard deviation evaluated near the centre of the observed data range, at $\Sigma_{\rm gas} = 10^{2.5}$~$M_\odot$~pc$^{-2}$ and $t_{\rm ff} = 1$~Myr. 
The entries for $\epsilon_{\rm ff}$ correspond to the mean and standard deviation of $\epsilon_{\rm ff}$ for each cloud (see Figure~\ref{fig:freefall}).
}
\end{rotatetable}
\end{table}

\clearpage
\bibliographystyle{aasjournal}
\bibliography{mybibliography}

\end{document}